\documentclass[10pt]{article}
\usepackage{ametsoc2col}

\newcommand{\myabstract}{Equilibrium statistical mechanics of two-dimensional flows provides an explanation and a prediction for the self-organization of large scale coherent structures. This theory is applied in this paper to the description of oceanic rings and jets, in the framework of a $1.5$ layer quasi-geostrophic model. The theory predicts the spontaneous formation of regions where the potential vorticity is homogenized, with strong and localized  jets at their interface. Mesoscale  rings are shown to be close to a statistical equilibrium: the theory accounts for their shape, their drift, and their ubiquity in the ocean, independently of the underlying generation mechanism. At basin scale, inertial states presenting mid basin eastward jets (and then different from the classical Fofonoff solution) are described as marginally unstable states. In that case, considering a purely inertial limit is a first step toward more comprehensive out of equilibrium studies that would take into account other essential aspects, such as wind forcing.
}

\begin{document}
\title{\textbf{\large{Oceanic rings and jets as statistical equilibrium states}}}
%
% Author names, with corresponding author information. 
% [Update and move the \thanks{...} block as appropriate.]
%
\author{\textsc{Antoine Venaille} \thanks{\textit{E-mail: venaille@princeton.edu}}\\
\textit{\footnotesize{GFDL-AOS, Princeton, O8540 NJ, US}}
\and 
\centerline{\textsc{Freddy Bouchet} \thanks{\textit{E-mail: freddy.bouchet@ens-lyon.fr}}}\\% Add additional authors, different institution
\centerline{\textit{\footnotesize{CNRS, ENS-Lyon, 69007 Lyon, France}}}
}

\ifthenelse{\boolean{dc}}
{
\twocolumn[
\begin{@twocolumnfalse}
\amstitle

\begin{center}
\begin{minipage}{13.0cm}
\begin{abstract}
	\myabstract
	\newline
	\begin{center}
		\rule{38mm}{0.2mm}
	\end{center}
\end{abstract}
\end{minipage}
\end{center}
\end{@twocolumnfalse}
]
}
{
\amstitle
\begin{abstract}
\myabstract
\end{abstract}
}

\newpage
 
\section{Introduction}

Large scale coherent structures are ubiquitous in the ocean. Understanding the physical mechanism underlying their formation and persistence  remains a major theoretical challenge. 

At mesoscale, oceanic  turbulence is mostly organized into westward propagating  rings, as for instance revealed by altimetry \citep{Chelton07}. Since typical eddy turnover times  are much shorter than dissipation and forcing time scales, these rings can be studied in the inertial limit, for which forcing and dissipation are neglected.

At basin scale, the dynamics are strongly influenced by  forcing and dissipation: wind forcing plays the leading role in setting the gyre structures through the Sverdrup balance, and  the concomitant effect of planetary vorticity gradients and dissipation explains their westward intensification \citep{Pedlosky:1998_OceanCirculationTheory}. Because none of these mechanisms  are conservative processes, the inertial approach does not take these essential aspects into account. Conversely, existing theories give no clear explanation of the existence of strong and robust eastward jets in the inertial part of these currents. The classical wind driven ocean theory and the inertial approach both give an incomplete picture, and complement each other. A useful step towards a  comprehensive non-equilibrium theory that would combine both approaches is to study mid-basin eastward jets in the inertial limit. Such is the focus of this paper. 

On the one hand, the problem of the self-organization of a turbulent flow involves a huge number of degrees of freedom coupled together via complex non-linear interactions. This situation makes any deterministic approach illusory, if not impossible. On the other hand, there can be abrupt and drastic changes in the large scale flow structure when varying a single parameter such as the energy of the flow. It is then appealing  to study this problem with a statistical mechanics approach, which reduces the problem of large-scale organization of the flow to the study of states depending on a few key parameters only. 

Such a theory exists: this is the Robert-Sommeria-Miller (RSM hereafter) equilibrium statistical mechanics  \citep{Robert:1990_CRAS,Miller:1990_PRL_Meca_Stat,Robert:1991_JSP_Meca_Stat,SommeriaRobert:1991_JFM_meca_Stat}. From the knowledge of the energy and the global distribution of potential vorticity levels provided by an initial condition, the theory predicts the large scale flow as the most probable outcome of turbulent mixing.

Here we ask the following question: can rings and jets be interpreted as RSM statistical equilibria in the framework of an $1.5$ layer quasi-geostrophic (QG) model?\\

The first attempt to use equilibrium statistical mechanics ideas to explain the self-organization of 2D turbulence was performed by  \cite{Onsager:1949_Meca_Stat_Points_Vortex} in the framework of the point vortex model. In order to treat flows with continuous vorticity fields, another approach has been proposed by Kraichnan in the framework of the truncated Euler equations \citep{Kraichnan_Motgommery_1980_Reports_Progress_Physics}, which has then been applied to  quasi-geostrophic flows over topography \citep{SalmonHollowayHendershott:1976_JFM_stat_mech_QG,Carnevale_Frederiksen_NLstab_statmech_topog_1987JFM}. The truncation has a drastic consequence: only the energy and the enstrophy  are conserved quantities, while any function of the  vorticity is conserved for the Euler equation. The energy-enstrophy statistical theory predicts the emergence of large scale mean flows above topography, characterized by a linear relationship between streamfunction and potential vorticity.

The existence of such a linear relation was assumed by \cite{Fofonoff:1954_steady_flow_frictionless}  for analytical convenience, in earlier work on inertial ocean circulation, independently of statistical mechanics approaches. Fofonoff was able to compute explicitly an inertial solution in the low energy limit. The emergence of such flows has then been observed in numerical simulations of freely evolving quasi-geostrophic flows \citep{ZouHolloway,WangVallis}. The energy-enstrophy statistical theory has therefore been proven successful to interpret  these Fofonoff flows as statistical equilibria. However, Fofonoff flows are not observed in the real ocean. 

There exists actually a richer variety of energy-enstrophy states, as revealed by the computations of solutions in various configuration, see e.g. \cite{Majda_Wang_Book_Geophysique_Stat,FrederiksenOKane08} and references therein. The computation of any equilibrium state characterized by a linear q-psi relation for a given energy and circulation (which includes the previous solutions, but not only) has been performed recently  for a large class of models including the $1.5$ layer QG equations \citep{Venaille_Bouchet_PRL_2009,AVFB}. None of the observed inertial features of oceanic flows, such as coherent rings and mid-basin eastward jets, were found in this class of equilibrium states.\\

For the Euler or QG dynamics, a major drawback of energy-enstrophy statistical theories is the loss of the additional invariants of the dynamics. A first attempt to include the effect of higher order invariant was proposed by \cite{Carnevale_Frederiksen_NLstab_statmech_topog_1987JFM}. The generalization of Onsager's ideas to the Euler and quasi-geostrophic equations with continuous vorticity field, taking into account all invariants, has lead to the RSM theory  \citep{Robert:1990_CRAS,Miller:1990_PRL_Meca_Stat,Robert:1991_JSP_Meca_Stat,SommeriaRobert:1991_JFM_meca_Stat}, see \cite{Eyink_Sreenivasan_2006_Rev_Modern_Physics,Majda_Wang_Book_Geophysique_Stat,Marston} for recent reviews and references. 

An essential point of the RSM approach is that it makes
a distinction between fine-grained potential vorticity distribution
on one hand, which is conserved by inviscid flows, and coarse-grained
distribution of potential vorticity on the other hand, which is not
conserved. This is in complete agreement with the classical and well
documented fact that potential vorticity invariants cascade towards smaller
and smaller scales (both for inviscid and dissipative flows), see e.g. \cite{Chavanis06}. Actually the equilibrium statistical mechanics predicts the ratio of potential vorticity invariants that remains to the large scale and the ratio
that cascade to smaller and smaller scales for an inviscid dynamics.

Computing the RSM equilibrium states requires the resolution of a variational problem with an infinite number of constraints. This practical difficulty has been overcome by treating canonically other invariants than the energy, which leads to a variational problem easier to solve \citep{Ellis00}. This approach is sometimes referred to as statistical mechanics with prior vorticity distribution \citep{Turk99,EHT,Majda_Wang_Book_Geophysique_Stat,Chavanis06}.  Any equilibrium state with prior vorticity distribution can be interpreted as an RSM equilibrium state \citep{Bouchet:2008_Physica_D}. This is therefore a very useful trick to compute the equilibrium states. However, we think that a physical interpretation of the prior vorticity distribution is problematic, since it would require to define what is a bath of potential vorticity. In the case of an isolated system (for instance a freely evolving inviscid flow), the relevant ensemble to consider is the one taking into account all the constraints of the dynamics. For this reason, we ultimately interpret the equilibrium states in term of the RSM theory.

Several studies, among which \cite{Abramov_Majda_2003_PNAS,Dubinkina_Frank_2010JCoPh}, have specifically addressed the importance of higher potential vorticity moments for the equilibrium states. Taking into account these additional dynamical invariants provides a much richer variety of equilibrium states than previous energy-enstrophy theories. It has been proven useful to describe the stratospheric polar vortex \citep{PrietoSchubert01}, the self-organization following deep convection events  \citep{DibattistaMajda00}, or Jupiter Red Great Spot \citep{Bouchet_Sommeria:2002_JFM,TurkingtonMHD:2001_PNAS_GRS}.\\

Most of oceanic coherent structures are surface intensified, with most of their kinetic energy located above the thermocline. In addition, eastward jets and rings are characterized by a jet width of the order of the first baroclinic Rossby radius of deformation. In this paper, we consider the simplest ocean model that takes into account this vertical structure and this typical horizontal length scale, namely an equivalent barotropic, $1.5$ layer QG model.

We consider the limit of small Rossby radius of deformation, which allows analytical computations of statistical equilibria, following the work of \cite{Bouchet_Sommeria:2002_JFM}. This assumption provides important insights for more general situations, even when such a scale separation does not exist. This is also a first step before considering the shallow water model, which is consistent with the scale separation between the Rossby radius of deformation and the domain scale.

In the limit of small Rossby radius of deformation, it has been shown that the computation of RSM statistical equilibria can be simplified into a Van-der-Waals Cahn-Hilliard variational problem \citep{Bouchet:2008_Physica_D}. These variational problems explain the formation and the shape of bubbles in thermodynamics. The existence of this formal analogy has been very fruitful in the description of Jovian vortices. This paper puts forward this approach in the oceanic context.  

The paper is organized as follows. Equilibrium statistical mechanics of the  $1.5$ layer QG model is presented in the second section. The method to compute analytically statistical equilibrium states in the limit of small Rossby radius of deformation is presented on the third section.  It allows for a justification of  the potential vorticity homogenization theory of \cite{RhinesYoung82}  without invoking any dissipation mechanism. The application to oceanic rings is discussed in a fourth section, by considering the case of a zonal channel on a beta plane. The application to mid-basin eastward jets  is discussed in a fifth section, by considering the case of a closed domain. Notations and symbols are referenced in table \ref{tab:notations}.

\section{Statistical mechanics of the $1.5$ layer, equivalent barotropic  QG model \label{sec:2D-Geostrophic-Turbulence}}

\subsection{The $1.5$ layer QG model, its dynamical invariants and its dynamical equilibria}

The simplest possible inertial mid-latitude ocean model taking into account the stratification of the oceans and the sphericity of the Earth is considered in this paper. This is the unforced, undissipated, $1.5$ layer QG model on a beta plane:
\begin{equation}
\frac{\partial q}{\partial t}+\mathbf{v}\cdot\nabla q=0, \quad \mathrm{with}\,\,\,\mathbf{v}=\mathbf{e}_{z}\times\nabla\psi  \ , \label{QG}\end{equation}
 \begin{equation}
\mathrm{and}\,\,\, q=\nabla^2 \psi-\frac{\psi}{R^{2}}+\beta y \ .\label{dir}\end{equation}
For the boundary conditions, two cases will be distinguished, depending on the domain geometry $\mathcal{D}$. In the case of a closed domain,  there is an impermeability constraint (no flow across the boundary), which amounts to a constant streamfunction along the boundary. To simplify the presentation, the condition  $\psi=0$ at boundaries will be considered\footnote{The physically relevant boundary condition should be  $\psi=\psi_{fr}$  where $\psi_{fr}$ is determined by using the mass conservation constraint $\int \mathrm{d} \mathbf{r} \ \psi= 0$ ($\psi$ is proportional to interface variations). Taking $\psi=0$ does not change  quantitatively the solutions in the domain bulk, but only the strength of boundary jets.}. In the case of a zonal channel, the streamfunction $\psi$ is periodic in the $x$ direction, and the impermeability constraint applies on northern and southern boundaries. In the remaining, length scales are nondimensionalized such that the domain area $|\mathcal{D}|$ is equal to one.\\

According to Noether's Theorem, each symmetry of the system is associated with the existence of a dynamical invariant, see e.g. \cite{Salmon_1998_Book}. These invariants are crucial quantities, because they provide strong constraints for the flow evolution. Starting from (\ref{QG}), (\ref{dir}) and the aforementioned boundary conditions  one can prove that QG flows conserve the energy: 
\begin{equation}
E=\frac{1}{2}\int_{\mathcal{D}}\mathrm{d}\mathbf{r}\,\left[(\nabla\psi)^{2}+\frac{\psi^{2}}{R^{2}}\right]=-\frac{1}{2}\int_{\mathcal{D}}\mathrm{d}\mathbf{r}\,\left(q-\beta y \right)\psi.\label{ene}
\end{equation}

Additionally, the QG dynamics (\ref{QG}) is a transport by an incompressible flow, so that the area $\gamma\left(\sigma\right) d\sigma$ occupied by a given vorticity level $\sigma$ is a dynamical invariant. The quantity $\gamma(\sigma)$ will be referred to as the global distribution of potential vorticity. The conservation of the distribution $\gamma\left(\sigma\right)$
is equivalent to the conservation of any moment of the potential vorticity $\int_{\mathcal{D}}\mathrm{d}\mathbf{r}\, q^n $, and is related to particle relabelling symmetry \cite{Ripa81, Salmon_1998_Book}.\\

The stationary points of the QG equations (\ref{QG}), referred to as \textit{dynamical equilibrium states}, satisfy $\mathbf{v} \cdot \nabla q=  \nabla \psi \times \nabla q =0$. It means that dynamical equilibria are flows for which streamlines are isolines of potential vorticity. Then, any state characterized by a $q-\psi$ functional relationship is a dynamical equilibrium.

At this point, we need a theory i) to support the idea that the freely evolving flow dynamics will effectively self-organize into a dynamical equilibrium state ii) to determine the $q-\psi$ relationship associated with this dynamical equilibrium iii) to select the dynamical equilibria that are likely to be observed. This is the goal and the achievement of equilibrium statistical mechanics theory,  presented in the next subsection.

\begin{table*}[t!]
\begin{center}

\caption{\label{tab:notations} Symbols  and notations used in the text.}

\begin{tabular}{ll}

\hline \hline
 Symbol  & definition \\
\hline
$ \mathbf{e}_{x,y,z} $&   unit vectors in the meridional ($x$), zonal ($y$) and vertical ($z$) direction\\
$\mathbf{r}=(x,y)$& coordinate of a point, with $\nabla=(\partial_x,\ \partial_y)$\\
$t$ & time coordinate\\
$R$ & Rossby radius of deformation\\
$f_0$  &  Coriolis parameter \\ 
 $\beta$ & planetary vorticity gradient \\
$\mathcal{D}$ & domain where the flow takes place (with $|\mathcal{D}|=1$)\\ 
$\mathbf{u}(\mathbf{r})$ & velocity field\\
$q(\mathbf{r},t)$& (fine grained) potential vorticity field\\
$\sigma$ & level of potential vorticity, with $\sigma\in \Sigma=]-\infty, \ +\infty [$ \\
$\overline{q}$, $\overline{\psi}$ & (coarse grained) mean field  potential vorticity and streamfunction\\
$E$ a& Energy of the flow \\  
$\gamma(\sigma)$& global distribution of potential vorticity levels \\
\hline
$\rho(\mathbf{r},\sigma)$& probability distribution function\\
$\mathcal{S}[\rho]$ & Mixing entropy\\
$\mathcal{N}[\rho]=N$ &  normalization constraint \\
$D_\sigma[\rho]=\gamma(\sigma)$  & constraint on global potential vorticity distribution\\
$\mathcal{E}[\rho]=E$ & constraint on the energy\\
$\zeta(\mathbf{r})$ &  Lagrange multiplier associated with the normalization $\mathcal{N}$\\
$\alpha(\sigma)$ & Lagrange multiplier associated with the global vorticity distribution $D_{\sigma}$\\
$\lambda$ & Lagrange multiplier associated with the energy $\mathcal{E}$\\
$g_{\alpha}(\lambda \psi)$ & $q-\psi$ relation at equilibrium \\ 
\hline
$\phi=\psi/R^2$ & rescaled streamfunction\\
$\mathcal{F}[\phi]$, $F$  & Free energy functional and equilibrium free energy \\ 
$f(\phi)$& specific free energy \\
$C=-\lambda R^2$& rescaled Lagrange multiplier associated with the energy constraint\\
$M=\int_\mathcal{D} \mathrm{d} \mathbf{r}\phi$ & constraint for $\phi$\\
$\mu$ & Lagrange parameter associated with  $M$\\
$\phi_{1}$, $\phi_2$ & values of $\phi$ in a given phase\\ 
$q_1$, $q_2$ & values of $q$ in a given phase \\
$A_1$, $A_2$ & domain (and area) occupied by a given phase\\
$L$ & perimeter of the interface between phases\\
$r$ & curvature radius of the interface\\
$\eta$ & Lagrange multiplier associated with the constraint on $A_2$  when minimizing $L$\\
$ \tau= R \widetilde{\tau}$ & coordinate across the interface \\
$\phi_{jet}(\tau)$ & jet profile\\
$U_{jet}=(\phi_2-\phi_1) R$ & velocity of the jet \\
$\mathcal{F}_{int}= c R L (\phi_2-\phi_1)^2$  & Free energy of the interface, with $c \sim 1$\\
\hline
$\mathcal{L}[q]=\mathcal{L}^i=\mathcal{L}^f$ & constraint on the linear momentum ($i$ : initial ; $f$ : final )\\
$y_f$, $y_{jet}$& latitude of the ring center of of jet latitude\\
$\widetilde{\beta}=\beta/R^2$& rescaled beta coefficient \\
$\mathcal{F}_{\beta}$ & contribution of the beta term to the free energy\\ 
$l(x)$ & perturbation of the zonal interface\\
$\mathcal{F}_R= \mathcal{F}_{int} + \mathcal{F}_{\beta}$& First order corrections to the free energy\\
$L_x$, $L_y$& zonal and meridional extension of the closed domain\\
$L_{ring}$& diameter of the ring\\
\hline

\end{tabular}
\end{center}
\end{table*}

\subsection{The equilibrium statistical mechanics of Robert-Sommeria-Miller (RSM) \label{sec:Equilibrium-statistical-mechanics}}

The RSM statistical theory is introduced on a heuristic level in the following. There exists rigorous justifications of the theory, see e.g. \cite{BouchetCorvellec10} and references therein. % \citep{Michel_Robert_LargeDeviations1994CMaPh.159..195M,Boucher_Ellis_Turkington_2000_JSP,Robert_2000_CommMathPhys-TruncationEuler,2004CMaPh.244..187E,BouchetCorvellec10}. 

A microscopic state is defined by its potential vorticity field $q(\mathbf{r})$. If taken as an initial condition, such a fine grained field would evolve toward a state with filamentation at smaller and smaller scales, while keeping in general a well defined large scale organization. Then, among all the possible fine grained states, an overwhelming number are characterized by these complicated small scale filamentary structures. This phenomenology gives a strong incentive for a mean-field approach, in which the flow is described at a coarse-grained level.

For that purpose, and following \cite{SommeriaRobert:1991_JFM_meca_Stat}, we introduce the probability $\rho(\sigma,\mathbf{r}) \mathrm{d}\sigma$  to measure a potential vorticity level $\sigma$ at a point $\mathbf{r}=\left(x,y\right)$. The probability density field $\rho$ defines a macroscopic state of the system. The corresponding averaged potential vorticity field, also referred to as \emph{coarse-grained}, or \emph{mean-field}, is 
\begin{equation} 
\overline{q}\left(\mathbf{r}\right)=\int_{\Sigma}  \mathrm{d}\sigma\,\,\sigma\rho\left(\sigma,\mathbf{r}\right) \ ,
\label{eq:qbar}
\end{equation} 
with the average streamfunction $\bar{\psi}$ defined by $\overline{q}=\nabla^2 \bar{\psi}-\bar{\psi}/R^{2}+\beta y  $, and where $\Sigma=]-\infty,\ +\infty [$. Many microscopic states $q$ can be associated with a given macroscopic state $rho$. The cornerstone of the RSM statistical theory is the computation of the most probable state $\rho_{eq}$, that maximizes the mixing  entropy given by the Boltzmann-Gibbs formula 
\begin{equation}
\mathcal{S}\left[\rho\right]\equiv-\int_{\mathcal{D}}\mathrm{d} \mathbf{r}\int_{\Sigma} \mathrm{d} \sigma\,\rho\log\rho \ ,
\label{eq:Entropie_Maxwell_Boltzmann}
\end{equation}
while satisfying the constraints associated with each dynamical invariant. The mixing entropy (\ref{eq:Entropie_Maxwell_Boltzmann}) is a quantification of the number of microscopic states $q$ corresponding to a given macroscopic state $\rho$. The state $\rho_{eq}$ is not only the most probable one: an overwhelming  number of microstates are effectively concentrated close to it \citep{Michel_Robert_LargeDeviations1994CMaPh.159..195M}. This gives the physical explanation and the prediction of the large scale organization of the flow.

In the remaining of this paper, the term \textit{global entropy maximum} or \textit{stable equilibrium state}  will be used for any global maximizer $\rho$ of the entropy (\ref{eq:Entropie_Maxwell_Boltzmann}) satisfying the constraints. The term  \textit{local entropy maximum} or \textit{metastable equilibrium state} will be used  for any state $\rho$ that is a  local maximizer of the entropy (\ref{eq:Entropie_Maxwell_Boltzmann}), satisfying the constraints.

To compute statistical equilibria, the constraints must be expressed in term of the macroscopic state $\rho$:
\begin{itemize}
\item The local normalization $ N\left[\rho\right](\mathbf{r})\equiv\int_{\Sigma} \mathrm{d}\sigma\,\,\rho\left(\sigma,\mathbf{r}\right)=1$,
\item The global potential vorticity distribution $D_{\sigma}\left[\rho\right] \equiv \int_{\mathcal{D}}d\mathbf{r}\,\rho\left(\sigma,{ \mathbf{r}}\right)=\gamma\left(\sigma\right) $,
\item The energy  $\mathcal{E}\left[\rho\right]\equiv -\frac{1}{2}\int_{\mathcal{D}}     \mathrm{d}\mathbf{r}  \int_{\Sigma}  \mathrm{d} \sigma \,  \rho \left({\sigma} - \beta y \right) \overline{\psi}  =E$.
\end{itemize}

Because of the overwhelming number of states with only small scale fluctuations around the mean field potential vorticity, and because energy is a large scale quantity, contributions of these fluctuations to the total energy are negligible with respect to the mean-field energy \cite{SommeriaRobert:1991_JFM_meca_Stat}. 

The first step toward computations of RSM equilibria is to find critical points  $\rho$ of the mixing entropy (\ref{eq:Entropie_Maxwell_Boltzmann}). In order to take into account the constraints, one needs to introduce the Lagrange multipliers $\zeta(\mathbf{r})$,  $\alpha(\sigma)$, and $\lambda$ associated respectively with  the local normalization,  the conservation of the global vorticity distribution and of the energy. Critical points are solutions of:
\begin{equation}
\forall\ \delta\rho\quad\delta\mathcal{S}-\lambda\delta\mathcal{E}-\int_{\Sigma} \mathrm{d}\sigma\ \alpha \delta D_{\sigma}-\int_{\mathcal{D}}\mathrm{d}\mathbf{r}\ \zeta \delta N=0 \ , \label{eq:critical_points}
\end{equation}
where first variations are taken with respect to $\rho$. This leads to $\rho =N \exp\left({\lambda\sigma\psi\left(\mathbf{r}\right)-\alpha(\sigma)}\right)$ where $N$ is determined by the normalization constraint $\left( \int \mathrm{d} \sigma\ \rho=1 \right)$. Finally, using (\ref{eq:qbar}), one finds that statistical equilibria are dynamical equilibria characterized by a functional relation  between  potential vorticity and  streamfunction:
\begin{equation}
\bar{q}=\frac{ \int_{\Sigma} \mathrm{d}\sigma\ \sigma e^{\lambda\sigma\psi\left(\mathbf{r}\right)-\alpha(\sigma)}}{  \int_{\Sigma} \mathrm{d}\sigma\,e^{\sigma \lambda\psi\left(\mathbf{r}\right)-\alpha\left(\sigma\right)} } \equiv g_{\alpha}\left( \lambda \bar{ \psi } \right)   \label{eq:q-psi_equilibre}
\end{equation}
It can be shown that $g_{\alpha}$ is a monotonic, increasing and bounded function of $\lambda \overline{\psi}$ for any global distribution $\gamma(\sigma)$ and energy $E$. These critical points can either be entropy minima, saddle or maxima. To find statistical equilibria, one needs then to select the entropy maxima.

At this point, two different approaches could be followed. The first one would be to consider a given small scale distribution $\gamma(\sigma)$ and energy $E$, and then to compute the statistical equilibria corresponding  to these parameters. In practice, especially in the geophysical context, one does not have empirically access to the fine grained vorticity distribution, but rather to the $q-\psi$ relation (\ref{eq:q-psi_equilibre}) of the large scale flow. The second approach, followed in the remaining of this paper, is to study statistical equilibria corresponding to a given class of $q-\psi$ relations.

More precisely, we will consider the class of $q-\psi$ relations that admit an inflexion point, referred to as ``$tanh$-like'' relations,  see figure \ref{fig:q-psi-relation}.  When the global distribution $\gamma(\sigma)$ is a double delta function (a two level system), one can explicitly show that the $q-\psi$ relation is a $tanh$ function \cite{Bouchet_Sommeria:2002_JFM}, but the actual class of initial conditions associated with $tanh$-like relations is expected to be much larger. A pragmatic point of view is that these $tanh$-like relations are the one that allow for statistical equilibria characterized by fronts of potential vorticity. In that respect, there are the relevant class of $q-\psi$ relations to describe either rings or zonal jets.

\subsection{Simplification of the variational problem \label{sub:QG strong jet generql}}

As explained in the introduction, a widely used method to solve a variational problems with many constraints (such as the RSM variational problem) is to consider a dual variational problem, which has the same critical points as the initial one, but that is less constrained: any solution of the less constrained (easier to solve) problem is a solution of the more constrained problem, see e.g.  \cite{Ellis00}.

Let us for instance consider the ensemble of rescaled streamfunction fields $\phi=\psi/R^2$, that satisfy the constraint $\int_{\mathcal{D}}\mathrm{d}{\mathbf{r}}\ \phi=M$. This constraint is equivalent to the conservation of the first moment of potential vorticity $\int_{\mathcal{D}}\mathrm{d}{\mathbf{r}}\ q$, at leading order in $R$. Let us then look for the minimizers of  the free energy functional

\begin{equation}
\mathcal{F}=\int_{\mathcal{D}}\mathrm{d}{\mathbf{r}}\,\left[\frac{R^{2}}{2} \left(\nabla\phi\right)^{2}+f\left(\phi\right)- \beta \phi y  \right] \ , \label{eq:Fphi}
\end{equation}
where $f(\phi)$ is a specific free energy to be defined precisely in the next paragraph. This provides a variational problem
\begin{equation}
F=\min_{\phi}\left\{ \mathcal{F}\left[\phi\right]\,\,\left|\,\, \int_{\mathcal{D}}\mathrm{d}{\mathbf{r}}\ \phi=M \right.\right\} \ , \label{eq:VdW-topo}
\end{equation}
which is much simpler than the one of the RSM statistical theory, since only one constraint is kept. Critical points of this problem are solutions of $\delta\mathcal{F}-\mu \int_{\mathcal{D}}\mathrm{d}{\mathbf{r}}\ \delta \phi =0$, for any perturbation  $\delta \phi$, where $\alpha$ is the Lagrange multiplier associated with the constraint. A part integration and  the relation $q=R^2 \nabla^2 \phi - \phi + \beta y$ give  $\delta\mathcal{F}=\int\mathbf{\mathrm{d}r}\ \left(f^{\prime}(\phi)-\phi-q\right)\delta\phi$. Critical points satisfy therefore the relation $q=f^{\prime}\left(\phi \right)-\phi-\mu$. These critical points are the same as the RSM critical points, given by equation (\ref{eq:q-psi_equilibre}), provided that 
\begin{equation}
f^{\prime}\left(\phi \right)=g_{\alpha}( \lambda R^2 \phi)+\phi +\mu \ . \label{eq:relation_f_g}
\end{equation}
This relation defines the specific free energy $f$. One can see that the $tanh$-like shape of $g_{\alpha}$ with a sufficiently steep slope at its inflection point  leads to a double-well shape for $f$, as illustrated on figure \ref{fig:q-psi-relation}. This double-well shape is an important ingredient for the computation of statistical equilibria in the following.

It has been proven by \cite{Bouchet:2008_Physica_D} that for each  minimizer $\phi$ of the variational problem (\ref{eq:VdW-topo}), there exists a set of constraints $E$, $\gamma(\sigma)$  such that $\phi$ is the  mean-field streamfunction of the RMS statistical equilibrium state $\rho$ associated with the constraints  $E$, $\gamma(\sigma)$. In other words, any local (global) free energy minimizer $\phi$ can be interpreted as a local (global) entropy maximum state of the RMS theory. 

\section{Potential vorticity homogenization and jets as statistical equilibria \label{sec:homogenization}}

Assuming that there is no beta effect ($\beta=0$), that $f(\phi)$ has a double-well shape, and considering the limit $R \ll L$ ($L$ is the domain size), \cite{Bouchet_These,Bouchet:2008_Physica_D} found that the variational problem (\ref{eq:VdW-topo}) becomes analogous to the Van der Waals -- Cahn -- Hilliard model  that describes phase separation and phase coexistence in thermodynamics \citep{1987_Modica_ArchRatMechAna}. This formal analogy provides then an interesting physical interpretation of self-organization phenomena in geostrophic turbulence.

\subsection{Solutions of the Van der Waals -- Cahn -- Hilliard variational problem \label{sub:Van Der Waals}}

\begin{figure}[t!]
\begin{center}
\includegraphics[width=0.7\columnwidth]{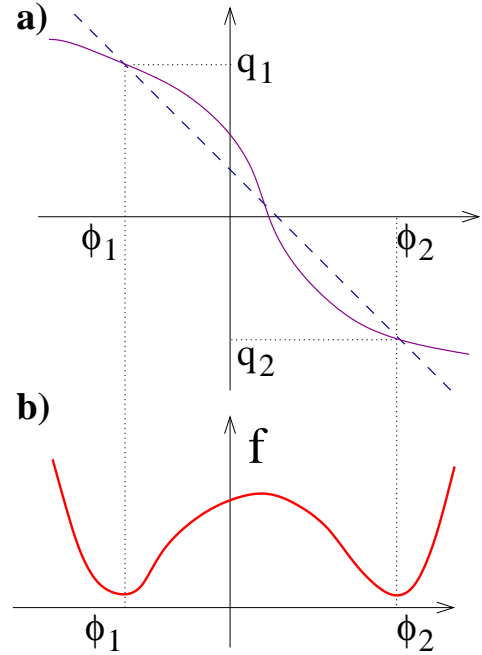}
\end{center}
\caption{{
\textbf{a)} In this paper, we consider the class of $q-\phi$  relations (with $\phi=  \phi/R^2$) having a $tanh$-like shape, namely i)decreasing with $\phi$ ii) bounded for $\psi \rightarrow\pm \infty$ iii) with a single inflection point. In addition, we assume that the slope at the inflection  point is sufficiently steep, so that the $q-\phi$ relation represented as a \textbf{plain line} cross three times the \textbf{dashed line} $q=-\phi-\mu$. \textbf{b)} The double-well shape of the specific free energy $f\left(\phi\right)$ appearing in the expression of the free energy functional (\ref{eq:Fphi}). This function is related to the $q-\psi$ relation through equation (\ref{eq:relation_f_g}). The Lagrange parameter $\mu$ is chosen such that the specific free energy of the two minima are the same ($f(\phi_1)=f(\phi_2)$), which allows phase coexistence.
}}
\label{fig:q-psi-relation}
\end{figure}

\begin{figure*}[t!]
\begin{centering}
\includegraphics[width=\textwidth]{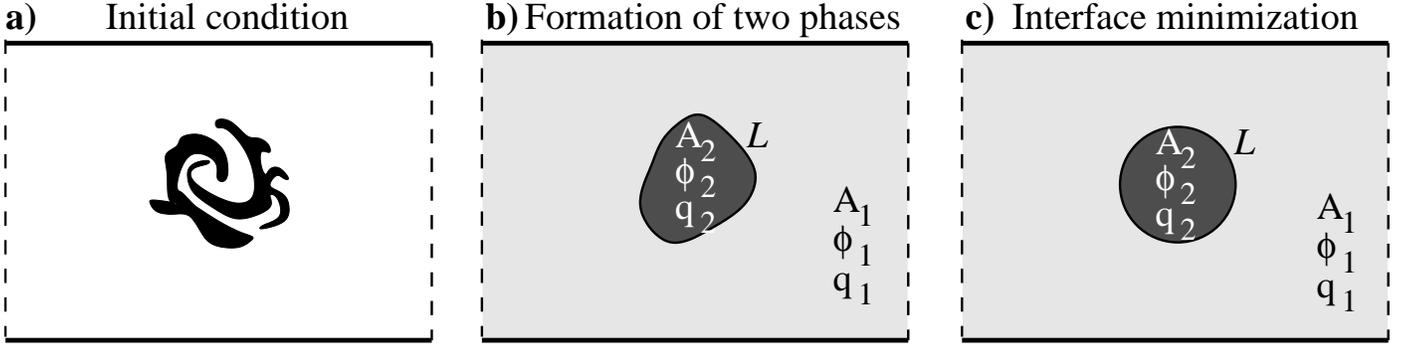}
\par\end{centering}

\caption{{Resolution of the Van der Waals variational problem (\ref{eq:VdW-topo}). a) Example of an initial condition for the potential vorticity field. Note that this initial condition could as well have many levels of potential vorticity. b) At zeroth order in $R$, $\phi$ takes  two values $\phi_1,\ \phi_2$ on two sub-domains $A_{1}$ and $A_2$ corresponding to the coexistence of two phases of homogenized potential vorticity $q_1$, $q_2$. These sub-domains are separated by strong jets of typical width $l_{jet} = R$ and velocity $U_{jet} = (\phi_2 -\phi_1) R$. c) The actual shape of the structure, or equivalently the
position of the jets, is obtained by minimizing its perimeter $L$ for a fixed value of $A_2$.}\label{fig_domaine}}
\end{figure*}

At zeroth order in  $R$, the function $f\left(\phi\right)$ plays the dominant role in the free energy functional $\mathcal{F}$  given by (\ref{eq:Fphi}). In order to minimize $\mathcal{F}$, the  streamfunction $\phi$  must therefore be equal to one of the two minima of the specific free energy $f(\phi)$ (the points $\phi_1$ and $\phi_2$ on figure \ref{fig:q-psi-relation}). Each of these minima corresponds to one phase. Without the constraint  $\int_\mathcal{D}\mathrm{d} \mathbf{r}  \phi =M$, one of the two uniform solutions $\phi=\phi_1$ or $\phi=\phi_2$ would  minimize $\mathcal{F}$: the system would have only one phase. But in order to satisfy the constraint $\int_\mathcal{D} \mathrm{d} \mathbf{r} \ \phi =M$, the system has to split into sub-domains: part of it with phase $\phi=\phi_1$ and part of it with phase $\phi=\phi_2$. In terms of free energy minimization, the coexistence of these two phases is possible  only if   $f(\phi_1)= f(\phi_2)$. Using equation (\ref{eq:relation_f_g}),  one can always choose the Lagrange parameter $\mu$ to satisfy this condition, see figure \ref{fig:q-psi-relation} for a graphical interpretation. In physical space, the area occupied by each of the phases is denoted $A_{1}$ and $A_{2}$ respectively, see figure \ref{fig_domaine}. These values are fixed by the constraint $\int_\mathcal{D} \mathrm{d} \mathbf{r} \ \phi =M$, which gives at leading order $\phi_1 A_{1}+\phi_2 A_{2}=M$ and by the geometrical constraint $A_{1}+A_{2}=1$ (where $1$ is the area of the domain).

The interface  between the sub-domains characterized by $\phi_1$ and $\phi_2$ corresponds to an abrupt variation of streamfunction. The term $R^{2}\left(\nabla\phi\right)^{2}$  in the expression (\ref{eq:Fphi}) of the free energy $\mathcal{F}$ is negligible everywhere except around this interface, on a typical width of order $R$. The interface is therefore associated with a strong and localized jet directed along this interface, with a typical velocity $U_{jet}=\left(\phi_1-\phi_2\right)R$, and a typical width $R$. The actual jet profile is computed in Appendix A, by minimizing the free energy associated with this profile. The jet gives always a positive contribution to the free energy:
\begin{equation}
\mathcal{F}_{int} = c \left(\phi_2-\phi_1\right) ^2  R L  \ , \ \ \text{with} \  c \sim 1  \ . \label{eq:Fint_final} 
\end{equation}

In order to minimize this interfacial free energy, the perimeter of the jet $L$ must be minimal, taking into account the constraints given by the fixed areas $A_{1}$ and $A_{2}$. We thus look for the curve with the minimal length that bounds a given surface. The solution of this classical problem is that the interface is either a circle or a straight line.

To conclude, the computation of the statistical equilibria predicts i) the formation of two phases of constant streamfunction, with strong and localized jets at these interface ii) the velocity profile across of these jets iii) the  shape of the interface, which is determined  by an isoperimetrical problem: the minimization of the interface length for a fixed enclosed area.

\subsection{Link with potential vorticity homogenization theories}

The $tanh$-like shape of the $q-\psi$ relations imply that sub-domains of constant streamfunction are also sub-domains of constant coarse grained potential vorticity. It means that the potential vorticity is homogenized in each sub-domain. Statistical mechanics provides therefore a physical explanation for the potential homogenization theory of \cite{RhinesYoung82}, without invoking any dissipation mechanism.

In the case of freely evolving 1.5 layer QG dynamics, statistical mechanics predicts not only the spontaneous formation of regions where potential vorticity is homogenized at a coarse grained level, but also the shape of the interface between these regions, corresponding to jets, where vorticity gradients are confined.

Statistical mechanics arguments also account for the irreversible nature of mixing: an overwhelming number of fine grained microscale are associated with a given coarse grained equilibrium state. The only effect of a weak small scale dissipation process would be to smooth out locally fine-grained fluctuations of potential vorticity, leaving unchanged its coarse-grained structure. 

Note that the formation of two subdomains of homogenized coarse-grained potential vorticity is essential to ensure the energy conservation. Importantly, any  eddy parameterization based on local down-gradient diffusivities could not represent this homogenization process: it would lead to the formation of a single phase of homogenized potential vorticity, which would in general not satisfy the energy constraint.  

These results are complementary to previous work focusing on the dynamics of potential vorticity mixing, using chaotic advection theory \citep{Pierrehumbert91}. Chaotic advection theory has been proven successful to account for many observed feature of potential vorticity mixing, but unlike statistical mechanics, it provides in general no prediction for the final state of the large scale flow, especially when there is no scale separation between mean and eddies in the initial condition. 

\section{Application to oceanic rings \label{sub:Gulf Stream Rings}}

Observations show that mesoscale oceanic rings  exist everywhere in the ocean, and particularly near western boundary currents, where high levels of eddy kinetic energy levels are largely associated with their presence,  see e.g. \cite{Olson91} for a review and  \cite{Morrow04,Chelton07} for recent altimetry measurements. Both cyclonic and anticyclonic rings propagate westward at speed $\beta R^2$. They also present an additional small meridional drift, poleward for cyclonic rings and equatorward for anti-cyclonic rings.

Among others, contributions of \cite{McWilliamsFlierlJPO79}, \cite{Nof81JPO}, of \cite{Flierl87} and of \cite{Cushman90} have provided insights on the dynamics and on the mechanisms responsible for the generation of the rings, in a large class of models and configurations. Various dynamical mechanisms accounting for the formation of the rings have been pointed out:  vortex shedding above topography, pinch off process during the non-linear evolution of a meandering jet, boundary layer separation, or large scale organization of initially small turbulent disturbances. Despite these very different generation mechanisms, the striking resemblance between observed oceanic rings in very different regions of the ocean, which has long been recognized as a \textit{surprising result} \citep{Olson91},  suggests that at least some aspect of these coherent structures can be studied independently of their generation mechanism. This is precisely the interest of statistical mechanics, which accounts  for spontaneous formation of circular structures surrounded by a jet, i.e. the self-organization of mesoscale turbulence into rings. 

\subsection{The westward drift of the rings \label{sub:The-westward-drift}}

To show that westward propagating circular rings can be interpreted as equilibrium states, two important ingredients must be taken into account: i) the beta effect $\beta y $  ii) a domain invariant by translation in the zonal direction. A drifting ring could not exist as statistical equilibria in a closed domain, since it would be destroyed when arriving on the western boundary. The zonal translational invariance of the problem has important consequences. It is shown in Appendix B that a change of Galilean reference frame in the zonal direction translates as a beta effect in the expression of potential vorticity. Moreover, in a reference frame moving at velocity $-\beta R^{2} \mathbf{e}_{x}$, the beta effect is exactly canceled out. We conclude that  in a domain invariant by translation in the zonal direction, statistical equilibria obtained  by the minimization of the Van-Der-Waals Cahn Hilliard variational problem (\ref{eq:VdW-topo}) without beta effect  are also statistical equilibria with beta effect, but drifting westward at speed $V=-\beta R^{2}$.

\subsection{The poleward drift of cyclones and the equatorward drift of anticyclones \label{sub:The-equatorward-drift}}

\begin{figure*}[t!]
\begin{center}
\includegraphics[width=\textwidth]{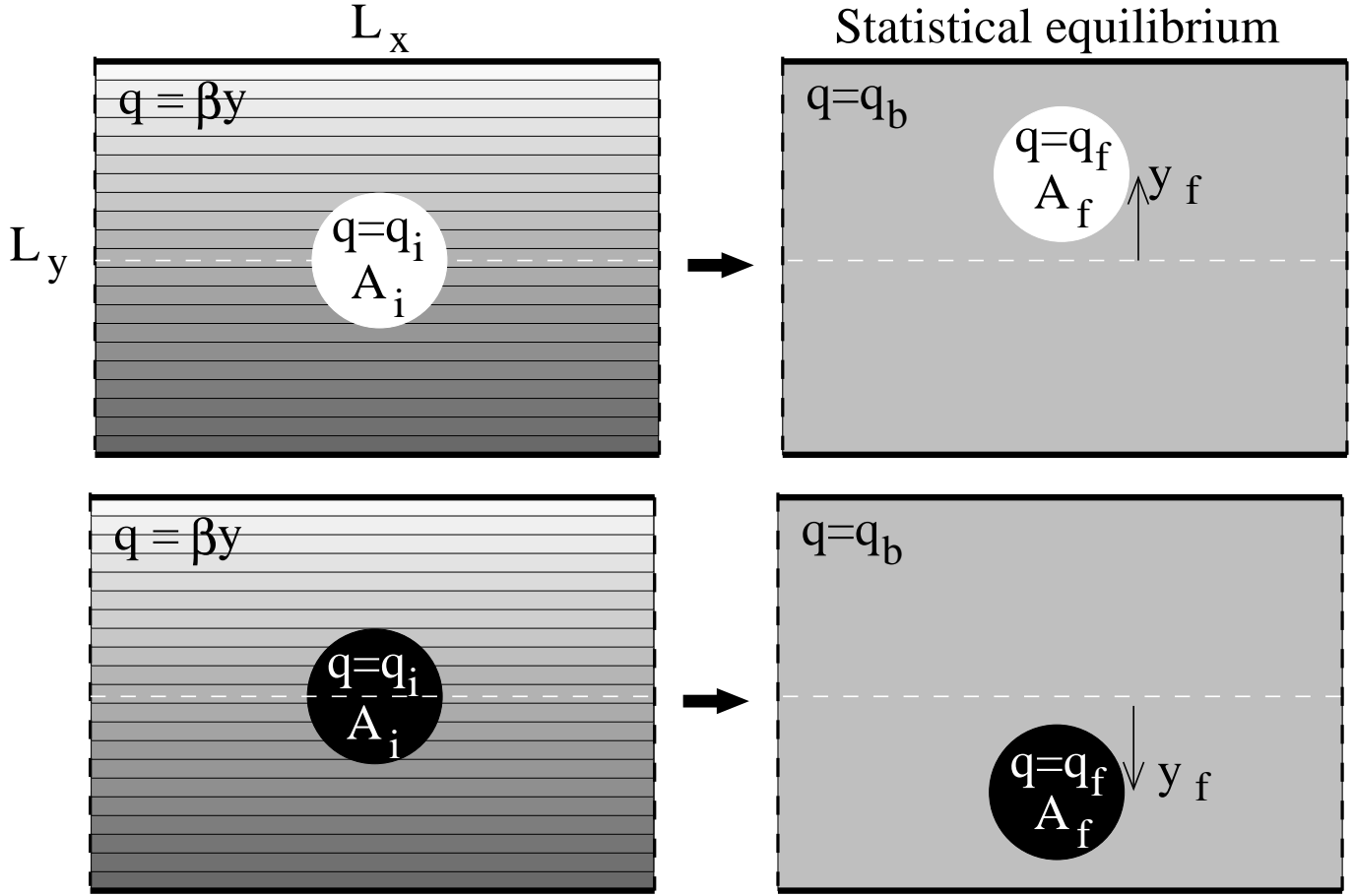}
\end{center}
\caption{Explanation of the meridional drift of rings, as a tendency to reach the statistical equilibria. On the left, initial conditions, with a white disk of positive potential vorticity in the upper panel, and a black disk of negative potential vorticity in the lower panel. In both cases the disk evolves on an initial beta plane with no background flow. On the right, the corresponding statistical equilibrium.}

\label{fig:drift}
\end{figure*}

If the flow actually reaches a local statistical equilibrium, then not only the ring is composed of an homogenized region of potential vorticity, but also the background flow. In figure \ref{fig:drift} is represented the case of an isolated patch of potential vorticity ($q=q_i$ within the ring of area $A_i$  centered on $y=0$) on a beta plane ($q=\beta y$ elsewhere). This situation is common in the ocean, for instance when Agulhas rings arrive in quiescent regions of the Atlantic ocean. Since the background potential vorticity is not homogenized, this state is  not a statistical equilibrium state.

It is shown in the following that the observed  asymmetric small meridional drift of cyclonic and anticyclonic rings can be understood as a tendency for the system to reach the statistical equilibrium (see also \cite{SchecterDubin01} for a similar argument in the context of beta plane turbulence). One needs for that purpose to consider the conservation of the linear momentum:
\[ \mathcal{L}=\int_{\mathcal{D}} \mathrm{d} \mathbf{r} \ q y \ ,\]
which is the dynamical invariant associated with the zonal translational symmetry. The linear momentum  of the initial condition is, at lowest order in $R$:
\[ \mathcal{L}^{i} \approx \beta \int_{\mathcal{D}}  \mathrm{d} \mathbf{r} \  y^{2}  -\beta \int_{A_{i}} \mathrm{d} \mathbf{r} \    y^{2}  \ ,\]
where $A_i$ is the initial area of the ring. Considering the limit of small rings compared to the domain size, the first term of the right hand side dominates the second one, and $\mathcal{L}^i \approx \beta L_x L_y^3/24 $. Assuming the statistical equilibrium is reached in the final state, the flow is made of two phases of homogenized potential vorticity: the background phase, with value $q=q_b$, and the ring's phase of area $A_f$, centered at latitude $y=y_f$,  with value $q=q_f$  (same sign as $q_i$) and with $|q_f|>|q_b|$.
The linear momentum of this final state is 
\[\mathcal{L}^{f}  \approx q_{b} \int_{\mathcal{D}}  \mathrm{d} \mathbf{r} \   y-\left(q_{f}-q_{b}\right)\int_{A_{f}} \mathrm{d} \mathbf{r} \   y \ .\]
Finally, the linear momentum conservation $\mathcal{L}^i=\mathcal{L}^f$ gives 
\[\frac{\beta}{24} L_x L_y^3\approx -\left(q_{f}-q_{b}\right) y_{f} A_{f}\ .\]
Physically, this means that for statistical mechanics reason, the background potential vorticity has to be homogenized, which leads to a loss of linear momentum that must be compensated by a latitude shift of the ring center. Rings with negative potential vorticity ($(q_f-q_b)<0$) decrease their latitude ($y_f<0$), while rings with positive potential vorticity $(q_f-q_b)>0$ increase their latitude ($y_f>0$). Then, in order to reach the statistical equilibrium, the ring has to drift  northward if it is made of an initial  positive potential vorticity patch, and southward if it is made of an initial negative potential vorticity patch. 
This corresponds always to a poleward drift for cyclonic structures and an equatorward drift for for anticyclonic structures, just as what is reported from altimetry measurements \citep{Morrow04,Chelton07}.

\subsection{Conclusion: oceanic rings are local statistical equilibria}

\begin{figure}[t!]
\begin{center}
\includegraphics[width=\columnwidth]{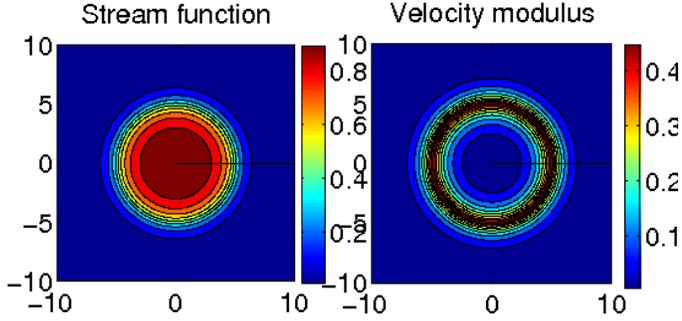}
\end{center}
\caption{Circular vortex as a statistical equilibrium of the QG model, with $R<L_{ring}$. Although analytical computations are carried in the limit $R\ll L_{ring}$,  the results are expected to hold  when this scale separation does not exit. It is a circular patch of (homogenized) potential vorticity in a background of homogenized potential vorticity, with two different values. The velocity field (right panel) has a ring structure.  The width of the jet surrounding the ring has the order of magnitude of the Rossby radius of deformation $R$.}

\label{fig:Rfini}
\end{figure}

In the ocean, the scale separation between the size of the rings and the Rossby radius of deformation is satisfied only to a limited extent. This scale separation has been assumed for technical reasons only: it allows for explicit analytical computations of the equilibria. The results obtained in this limit actually apply for far more general situations. This is confirmed by numerical computations of RSM equilibria, in which rings are obtained as local equilibria even when  the scale separation is not satisfied, see e.g. figure \ref{fig:Rfini}.

To conclude, the quasi-circular shape of oceanic rings and their westward propagation suggest that these coherent structures can be interpreted as local statistical equilibria. The existence of a meridional drift shows a departure from the prediction of the equilibrium theory. However, the fact that this drift can be interpreted as a tendency to reach to equilibrium state shows that these structures remain close to an equilibrium state.

Rings are local (metastable), and not global statistical equilibria of the equivalent barotropic model: a global equilibria would imply the coalescence of all existing rings into a single large scale vortex, in order to minimize the total interface between the different  regions of homogenized potential vorticity.

\section{Application to mid-basin eastward jets \label{sub:Gulf Stream and Kuroshio}}

Another region of the ocean where strong jets of typical width given by the Rossby radius of deformation $R$  are localized along an interface   separating two  regions of homogenized potential vorticity is the inertial part of mid-latitude eastward jets, as the Kuroshio or the Gulf stream currents. The inertial part of these currents is located in the regions where the western boundary currents separate from the coastline and self-organize downstream into a  strong eastward jet. Because mid-basin eastward jets fill a large part of oceanic basins, and because the existence of a western boundary is an essential ingredient for their formation, one must look for statistical equilibria in a close domain in that case. In view of the applications to mid-basin ocean jets, we assume a situation in which the global distribution of potential vorticity is symmetric: two phases characterized by symmetric values of streamfunction ($\phi_1=-\phi_2=U_{jet}/(2R)$), each of them filling half the domain area  ($A_1=A_2=1/2$). We ask in this section whether configurations with mid-basin eastward jets are statistical equilibria.

\subsection{Without beta effect, mid basin eastward jets are statistical equilibria of the QG model \label{sub:Eastward-jets-without_equivalent topography}}

\begin{figure*}[t]

\begin{centering}
\includegraphics[width=\textwidth]{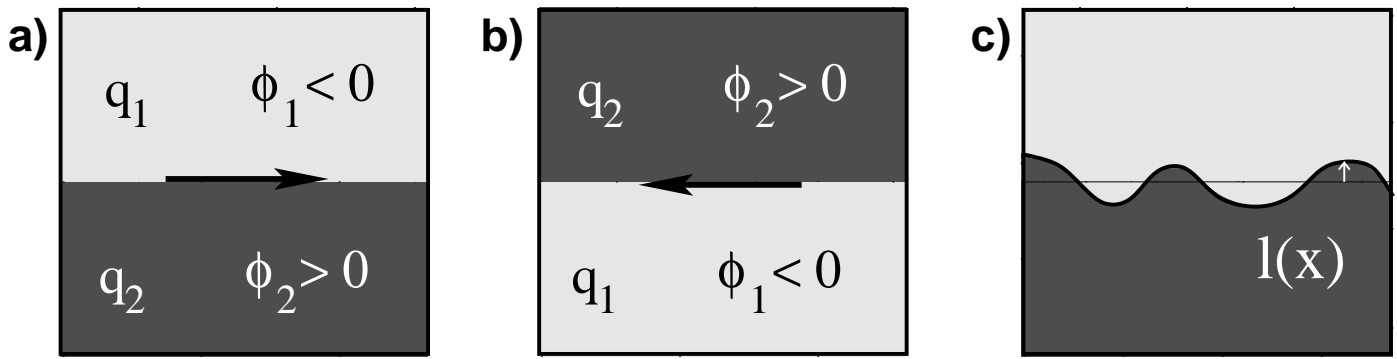}
\par\end{centering}
 
\caption{{ a) Eastward jet configuration b)
Westward jet configuration c) Perturbation of the
interface for the eastward jet configuration, to determine when this
solution is a local equilibrium. Without beta effect, both the eastward and the westward configurations are entropy maxima. With positive beta effect the westward jet becomes the the global entropy maximum, and the eastward jet becomes metastable provided that beta is small enough. 
\label{fig_pvfronts}
}}
\end{figure*}

The value $\phi=\phi_{1,2}$ for the two coexisting phases is not compatible
with the boundary condition $\phi=0$. As a consequence, there exists a boundary jet in order to match a uniform phase $\phi=\phi_{1,2}$ to the boundary conditions. Just like interior jets, treated in section \ref{sec:homogenization}, these jets contribute to the first order free energy, which gives the boundary jet structure and shape.  The symmetry of the problem ($\phi_1=-\phi_2$) implies that boundary jets of each phase give the same contribution to the free energy. Because the boundary length is a fixed quantity, the free energy minimization amounts to the minimization of the interior jet length only, just as in previous subsections.  The interior jet position and shape is thus given by the minimization of the interior jet length with fixed area $A_{2}=1/2$.

The jet has to be straight or circular. There are three possible interface configurations with straight or circular jets: i) the zonal jet configuration (jet along the $x$ axis) with $L=L_{x}$, ii) the meridional jet configuration (jet along the $y$ axis) with $L=L_{y}$, iii) and an interior circular vortex, with $L=\sqrt{2\pi}$
. The case i) of a zonal jet is a global interface minimum (and then a global equilibrium state) if and only if the aspect ratio satisfies $L_{x}/L_{y}<1$. For $L_{x}/L_{y}>1$, these solutions become metastable states (local entropy maximum). 

We conclude that without beta effect, mid-latitude eastward jets are statistical equilibria. Because of the symmetry  $\phi_1=-\phi_2$, solutions presenting eastward and westward jets are equivalent:  westward jets are also statistical equilibria. 

\subsection{With beta effect, eastward jets becomes metastable or unstable \label{sub:Eastward-jets-with_equivalent topography}}

\begin{figure}[t]
\begin{center}
\includegraphics[width=0.8\columnwidth]{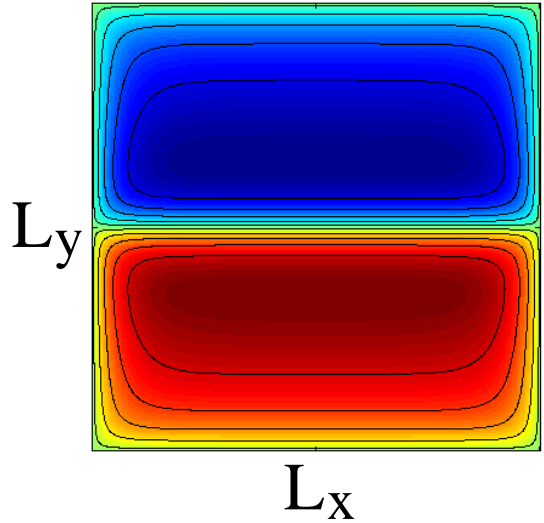}
\end{center}
\caption{{
Streamfunction of the solution presenting an eastward jet with beta effect (red: positive values, blue: negative values), associated with case  (a) of figure \ref{fig_pvfronts}.  The jet width is of order $R$. This solution is a statistical equilibrium for $L_x< \pi (U_{jet}/\beta)^{1/2}$.
}}
\label{fig:eastward-jet}
\end{figure}

Contrary to the case of the zonal channel, the beta effect can not be cancelled out by a change of Galilean reference frame in the case of a closed domain. One can therefore not avoid taking into account this term in the computation of the equilibrium free energy. One can readily see on the expression (\ref{eq:Fphi}) of the free energy, that when $\beta \ne 0$, the additional term $\mathcal{F}_{\beta} \equiv -\beta \int_\mathcal{D} \mathrm{d} \mathbf{r} \ \phi  y$ breaks the symmetry $\pm q $. The westward jet case  (with $\phi<0$ on the southern part of the domain and $\phi>0$ on the northern part) is more  favorable in terms of free energy minimization than the eastward jet case (with $\phi>0$ in the southern part of the domain and  $\phi<0$ in the northern part): westward jets become the only global equilibria for $\beta>0$ and aspect ratio $L_x/L_y>1$.

Let us be more precise by considering the limit of small beta effect, with the scaling $\beta \sim R \widetilde{\beta}$. With that scaling, the equivalent topography does not play any role at zeroth order in the variational problem (\ref{eq:VdW-topo}). We thus still conclude that phase separation occurs, with sub-domains of fixed areas $A_{1}$ and $A_{2}$, separated by jets whose transverse structure is described in Appendix A. It is shown in this same Appendix that the interface gives a contribution $\mathcal{F}_{int}=c R L \left( \phi_2-\phi_1 \right)^2 $.  Using the zeroth order result $\phi= \phi_1$ on sub-domain $A_1$ and $\phi=\phi_2$ on sub-domain $A_2$, one obtains also $\mathcal{F}_{\beta}=- \left(\phi_2 -\phi_1\right) \beta \int_{A_{2}}\mathrm{d}{ \mathbf{r}}\  y \ $, plus an unimportant constant. Finally, the total first order contribution to the free energy is
\begin{equation}
\mathcal{F}_R=c R \left( \phi_2-\phi_1 \right) ^2  L -  \left( \phi_2-\phi_1 \right) \beta \int_{A_{2}}\mathrm{d}{ \mathbf{r}} \ y \ . \label{Energy_libre_ordre1}
\end{equation}

Recalling that first variations of the length are proportional to the inverse of the curvature radius $r$ of the interface \citep{GelfandBook}, the minimization of (\ref{Energy_libre_ordre1}), with fixed area $A_{2}$ gives 
\begin{equation}
  \left( \phi_2-\phi_1 \right)  R \widetilde{\beta} y +\eta=\frac{c R  \left( \phi_2-\phi_1 \right) ^2}{r} \ , \label{eq:Rayon_courbure_h}
\end{equation}
where $\eta$ is a Lagrange parameter associated with the conservation
of the area $A_2$.

We conclude that zonal jets (i.e. an interface at latitude $y=y_{jet}$, with infinite curvature radius $r$) are
solutions to this equation for $\eta=- R \left(\phi_2 -\phi_1\right) \widetilde{\beta} y_{jet}$. This shows that eastward and westward jets described in the previous section are therefore still critical points of entropy maximization.\\

The eastward jet configuration is the one with the region $A_2$ below the line $y=0$ at the center of the domain. To determine if this configuration is a local statistical equilibria, let us consider  perturbations of this interface (given by the line $y=l(x)$), while keeping constant the area occupied by both phases, see figure \ref{fig_pvfronts}.  There are two contributions competing with each other in the expression (\ref{Energy_libre_ordre1}) of the free energy  $\mathcal{F}_R$. Any perturbation increases the jet length $L=\int \mathrm{d} x \ \sqrt{1+\left(l^{\prime}\right)^{2}}$, where $l^{\prime}=dl/dx$, and then increases the first term of the free  energy (\ref{Energy_libre_ordre1}) by $\delta \mathcal{F}_{int}=c R \left(\phi_2 -\phi_1\right)^2  \int dx\,\left(l^{\prime}\right)^{2}$. Any perturbation decreases the second term of the free energy (\ref{Energy_libre_ordre1}) by $\delta \mathcal{F}_{\beta}=-  R \left(\phi_2 -\phi_1\right) \widetilde{\beta} \int dx\,\ l^{2}$. If the eastward jet solution is not a free energy minimum, it exists a perturbation of the interface leading to negative variations of the free energy $\delta \mathcal{F}_R=\delta\mathcal{F}_{int}+\delta\mathcal{F}_{\beta}$. Let us consider the particular case $l=l_{k}\sin \left( k \pi x / L_{x} \right)$ where $k\ge1$ is an integer. Using  $U_{jet}=\left(\phi_2 -\phi_1\right) R$ and $\beta=R \widetilde{\beta}$, the condition $\delta \mathcal{F}_R <0$ gives $\beta>c  U_{jet} \left( k\pi /L_{x}\right)^{2}$. The most unfavorable case is for the smallest value of $k^{2}$, i.e. $k^{2}=1$.

It leads to the  necessary condition $ \beta > c U_{jet} \pi^{2} / L_{x}^{2}  $ for the eastward jet solution to be unstable (in term of statistical mechanics). It can actually be shown, using less straightforward considerations, that it is also a sufficient condition for instability. The \emph{destabilizing} effect of increasing values of $\beta$ contrasts with its stabilizing effect in classical criteria for barotropic instability, see e.g. \cite{VallisBook}. It has actually been shown that such eastward jet solutions can simultaneously be unstable for statistical mechanics, and stable for non-linear perturbations  \citep{Venaille_these,Venaille_Simonnet_Bouchet_2008_Eastward_Jets}.

For a fixed value of $\beta$,  eastward jets are local free energy maxima if the domain zonal extension is smaller than a critical value: $ L_{x}<\pi\left({c U_{jet}}/{\beta}\right)^{1/2}$. The streamfunction of such a state is presented in figure \ref{fig:eastward-jet}. For jets like the Gulf Stream,  $U_{jet} \approx 1 \ m.s^{-1}$ and $\beta \approx 10^{-11}\ m^{-1}.s^{-1}$. Using $c\sim 1$, the critical domain length scale upon which eastward jet become unstable is  $L_{x} \approx 300\ km$. This length is  smaller than the typical zonal extension of the inertial part of the Kuroshio or Gulf Stream currents, but not by an order of magnitude, which suggests that these structures are marginally unstable. The instability is consistent with the fact that strong meanders and pinch off process occurs downstream of oceanic eastward jets. But the marginal nature of this instability is also consistent with the overall robustness of the global structure of the flow, which becomes a statistical equilibrium when  the extension of the jet is small enough.

\section{Conclusion and prospects}

The aim of this paper was to present a point of view complementary to existing approaches that deal with coherent structures in the ocean. It was shown that the RSM statistical mechanics provides a unified framework that may be useful to study mesoscale and basin scale inertial flows.

The theory gives a physical explanation and a prediction for the self-organization of large scale oceanic coherent structure, independently of the underlying generation mechanism. It predicts the formation of subdomains of homogenized potential vorticity, with intense jets at the interface. Mesoscale rings can be interpreted as local equilibrium states of the RSM theory. Their shape and their drift can be understood in this framework. Mid-basin eastward jets are found  marginally unstable states of the RSM theory, consistently with observations of these jets.

The interest of this  approach relies on its generality (it does not depends on a particular flow configuration) and on its ability to describe qualitatively different observed regimes of self-organization, such as rings and zonal jets. The present study was achieved in the framework of  a $1.5$ layer QG model, which is too simplistic to describe oceanic eddies quantitatively; however, generalizations and further investigations in the framework of more complex models can be built upon these results.

A caveat of this approach is that forcing and dissipation are not taken into account in the framework of the  equilibrium theory: the input of the RSM theory is given by the dynamical invariants. In the case of mesoscale rings, even if the dynamics can be considered close to an equilibrium state,  forcing and dissipation play an important role in setting these dynamical invariants. In the case of basin scale jets, their marginal instability suggest that one can not avoid taking into account forcing and dissipation mechanisms to explain these structures. So far the inertial part of wind driven circulation has been mostly studied  from the point of view of bifurcation theory, starting from a highly dissipated ocean and  decreasing progressively frictional parameters, see e.g. \cite{DijkstraGhil05}. We argue that this problem can be tackled with another point of view, starting from the purely inertial limit (this paper), and adding small forcing and dissipation (future work built upon \cite{Bouchet_Simonnet_2009}). These two approaches are complementary and may be combined in the future in a more comprehensive non-equilibrium theory.
%%%%

\begin{acknowledgment}
It is a pleasure to thank Joel Sommeria for collaboration on statistical mechanics approach, as well as Stephen Griffies, Isaac Held and Geoffrey Vallis for interesting discussions that helped to improve the manuscript. The authors also warmly thank Nicolas Sauvage for his preliminary work during a traineeship with FB in 2005. This work was supported by the ANR program STATFLOW ( ANR-06-JCJC-0037-01 ) and the ANR program STATOCEAN (ANR-09-SYSC-014). AV was also supported by DoE grant DE-SC0005189 and NOAA grant NA08OAR4320752 during part of this work.
\end{acknowledgment}

\ifthenelse{\boolean{dc}}
{}
{\clearpage}
\begin{appendix}[A]
\section*{\begin{center} Computation of the jet profile \end{center}}

At leading order, minimization of the free energy leads to the formation of subdomains of constant streamfunction $\phi=\phi_1$ and $\phi=\phi_2$. The interface between these subdomains are associated with strong and localized jets.  Let us assume that the curvature radius of the interface is much larger than $R$, which allows us  to neglect what happens along the interface at leading order. Calling $\tau =R \widetilde{\tau}$ the coordinate in a direction along the normal to the interface, the jet profile  $\phi_{jet}(\tau)$ across the interface must be such that it minimizes its contribution to the total free energy  (\ref{eq:Fphi}).  The jet profile is therefore determined by solving a one dimensional variational problem:

\begin{equation}
\mathcal{F}_{int}=  L R \ \min_{\phi_{jet}} \left\{ \int_{-\infty}^{+\infty}\mathrm{d}\widetilde{\tau}\,\left[\frac{1}{2}\left(\frac{d\phi_{jet}}{d\widetilde{\tau}}\right)^{2}+f(\phi_{jet})\right]\right\} ,\label{eq:Variational Free Energy Unit lenght} 
\end{equation}

where $L$ is the perimeter of the jet and $\mathcal{F}_{int}$ the free energy associated with the existence of this interfacial jet. Critical points of this variational problem are states that cancel the first variations  of the free energy with respect to $\phi_{int}$. They are solutions of 
\begin{equation}
\frac{d^{2}\phi_{jet}}{d\widetilde{\tau}^{2}}=\frac{df}{d\phi_{jet}} \ .\label{jet}
\end{equation}

Making an analogy with  mechanics, if $\phi_{jet}$ would be a particle position, $\tau$ would be the time, equation (\ref{jet}) would describe the conservative
motion of the particle in a potential $-f$.
In order to connect the two different phases in the bulk, on each
side of the interface, one has to consider solutions with boundary
conditions $\phi\rightarrow \phi_1$ for $\tau \rightarrow - \infty$ and $\phi \rightarrow \phi_2$ for $\tau\rightarrow + \infty$. It exists a unique trajectory with such limit conditions. In the particle analogy, it is the trajectory connecting the two unstable fixed points $\phi_1$ and $\phi_2$, corresponding to the two bumps of the potential $-f$ (see figure \ref{fig:q-psi-relation}).

The energy  $\left(d\phi_{jet}/d\tau \right)^{2} /2-f(\phi_{jet})$ is conserved during the evolution of $\phi$ with time $\tau$. Using this conservation property and the boundary condition $ \phi \rightarrow \phi_2 $ for $\tau \rightarrow + \infty$, one obtains $f(\phi_{jet})=\left(d\phi_{jet}/d\tau \right)^{2} /2$, plus an unimportant constant. Injecting this expression into the variational problem (\ref{eq:Variational Free Energy Unit lenght}), one obtains 
\[\mathcal{F}_{int}=L R \int_{-\infty}^{+\infty}\left(d\phi_{jet}/ d\tau \right)^{2}d\tau = c  R L \left(\phi_2 -\phi_1 \right)^2 , \]
with $\ c\sim 1$. An important physical consequence is that the jet at the interface always gives a positive contribution to the free energy of the equilibrium state, which is of order  $R$ and proportional to the interface length $L$.
\end{appendix}

\ifthenelse{\boolean{dc}}
{}
{\clearpage}
\begin{appendix}[B]
\section*{\begin{center} Galilean invariance and beta effect \end{center}}
In the case of a zonal channel, the QG equations (\ref{QG}) are invariant over a Galilean transformation in the zonal direction 
\[x\rightarrow x^\prime=x-V t\ , \quad  y\rightarrow y^\prime=y \ , \quad  t\rightarrow t^\prime=t \ .\]
The velocity is transformed as $\mathbf{v} \rightarrow \mathbf{v'}=\mathbf{v}-V\mathbf{e}_{x}$, which, using the relation $\mathbf{v}=\mathbf{e}_{z} \times \nabla\psi$, gives the transformation law for the streamfunction $ \psi \rightarrow \psi'=\psi-Vy$. 
From the expression $q=\Delta\psi-\psi/R^{2}+\beta y$, one obtains finally  the transformation law for the potential vorticity $q\rightarrow q'=q+Vy/R^{2}$.
Thus the expression for the dynamics in the new reference frame is 

\[ \frac{\partial q^{\prime}}{\partial t^\prime}+\mathbf{v}^{\prime} \cdot \nabla^{\prime}q^{\prime}=0, \quad \mathrm{with}\,\,\,\mathbf{v}^{\prime}=\mathbf{e}_{z}\times\nabla^{\prime}\psi^{\prime}  \ ,\]

 \[
\text{and} \quad  q^{\prime}=\nabla^{\prime 2} \psi^{\prime}-\frac{\psi^{\prime}}{R^{2}}+\left(\beta +\frac{V}{R^{2}}\right)y^{\prime}\ .\]
\end{appendix}

\ifthenelse{\boolean{dc}}
{}
{\clearpage}

\bibliographystyle{ametsoc}

\bibliography{./bib/FBouchet,./bib/Meca_Stat_Euler,./bib/Ocean,./bib/Turbulence_2D,./bib/rings,./bib/Statistical-Mechanics,./bib/FBouchet-Proceedings,./bib/Jets-QuasiGeostrophic,./bib/FBouchet-Autres.bib}

\end{document}